\documentstyle{mn}
\def\go{
\mathrel{\raise.3ex\hbox{$>$}\mkern-14mu\lower0.6ex\hbox{$\sim$}}
}
\def\lo{
\mathrel{\raise.3ex\hbox{$<$}\mkern-14mu\lower0.6ex\hbox{$\sim$}}
}
\def\simeq{
\mathrel{\raise.3ex\hbox{$\sim$}\mkern-14mu\lower0.4ex\hbox{$-$}}
}

\def\ie{{\it i.e.\ }}

\def\ie{{\it i.e.\ }}

\def\be{\begin{equation}}
\def\ee{\end{equation}}
\def\bea{\begin{eqnarray}}
\def\eea{\end{eqnarray}}

\def\ie{{\sl i.e.\ }}

\def\hw2{{\hat W}^2}
\def\go{\mathrel{\raise.3ex\hbox{$>$}\mkern-14mu
             \lower0.6ex\hbox{$\sim$}}}
\def\lo{\mathrel{\raise.3ex\hbox{$<$}\mkern-14mu
             \lower0.6ex\hbox{$\sim$}}}
\def\ltorder{\mathrel{\raise.3ex\hbox{$<$}\mkern-14mu
             \lower0.6ex\hbox{$\sim$}}}
\def\gtorder{\mathrel{\raise.3ex\hbox{$>$}\mkern-14mu
             \lower0.6ex\hbox{$\sim$}}}

\def\eps2{{\epsilon^2}}

\def\msun{{\rm M_{\odot}}}

\input epsf.sty
\input psbox.tex

\title[Planets in 47 Tuc]
{Planets in 47 Tuc}
\author[Davies and Sigurdsson]{Melvyn B. Davies$^1$ and 
Steinn Sigurdsson$^2$ \\
$^1$ Department Physics \& Astronomy, University of Leicester, 
Leicester, LE1 7RH \\
$^2$ Astronomy Department, Pennsylvania State University, University Park,
PA 16802}

\begin{document}

\date{Received ** *** 2000; in original form 2000 *** **}

\label{firstpage}

\maketitle

\begin{abstract}
We consider the survivability of planetary systems in the globular cluster
47 Tucanae. We compute the cross sections for the breakup of planetary
systems via encounters with single stars and binaries. We also compute
the cross sections to leave planets on eccentric orbits. We find that
wider planetary systems ($d \go 0.3$ AU) are likely to be broken up in
the central regions of 47 Tucanae (within the half-mass
radius of the cluster). However tighter systems, and those in 
less-dense regions may survive. Tight systems will certainly survive
in less-dense clusters where subsequent surveys should be conducted. 
\end{abstract}


\section{Introduction}


The search for extra-solar planets came to fruition in 1992
when Alex Wolszczan (1992) discovered the first confirmed 
extra-solar planets orbiting the pulsar 1257+12.
Since Wolszczan's discovery, conventional optical searches have
found a number of extra-solar planets, biased towards the so-called
``hot Jovians'', typified by 51 Peg \cite{ma95,ma96}.
The observing techniques, measurement of low amplitude 
periodic Doppler shifts in the primary, bias such searches towards
high mass planets with short orbital periods.
Recently the nature of these systems, gaseous giant planets in short
period orbits, was confirmed by the observation of a transit in the
HD 209458 system \cite{ch00}.

A second pulsar-planet system has also been observed. 
PSR~B1620-26 in the globular cluster M4 is a binary pulsar
with a low mass white dwarf secondary in a $\sim 1$ AU near
circular orbit \cite{ly88}. In 1993 Backer et al discovered the
presence of a second orbiting companion, conjectured to be
have a planet mass \cite{ba93,si93,th93}. Subsequent observations
have confirmed the presence of a  second companion around 1620-26.
The second companion has sub-stellar mass and
it is near certain that the object is a planet \cite{th99}.
M4 is a modest, relatively low metallicity cluster, and the 
most plausible formation scenario requires that the planet
was exchanged into the pulsar system during an encounter
with a main--sequence star in the cluster \cite{si93,si95,jo97,fo00}.
The implication is that Jovian mass planets must exist around 
at least some of the main-sequence stars in relatively low metallicity
globular clusters.  The progenitor system of the M4 planet must have
had the planet in a wide orbit just before the exchange encounter 
($a \gg 1 AU$), in order for it
to have been plausibly exchanged into the current configuration.


Recently, a {\it Hubble Space Telescope} project (GO-8267) carried
out a transit search for ``hot Jovians'' in the globular cluster
47 Tucanae. The observations involved the continuous monitoring of
several tens of thousands of stars using WF/PC2 and STIS.
The basic technique was to take short (few minute) exposures,
primarily alternating I and V band. A single field in the cluster,
with the PC chip centered on the edge of the cluster core, was
monitored continuously for over 8 days.

The search is sensitive to Jovian sized planets with orbital periods
$\ltorder 4$ days, orbiting main-sequence stars in the cluster.
A detection requires two observations of a dimming of a few millimagnitudes
in both bands, lasting several hours (typical transit times scales are 3 hours).
Only relative photometry is required, and about 34,000 stars
were monitored simultaneously.

If the distribution of orbital parameters of Jovian planets in 47 Tuc
is comparable to that in the stellar neighbourhood, for short orbital periods,
then given a random orientation of orbital planes, the search should
discover some 15--20 51 Peg like planets.


Globular clusters are crowded environments. 
Binary systems undergo dynamical interactions with nearby stars, and
close encounters can lead to large changes in orbital parameters, including
the disruption of the system \cite{he75,hb83}. 
It is of interest to consider the probability of a planetary system surviving
for a time comparable to the Hubble time in the inner regions of globular cluster,
and the possible dynamical evolution of high mass ratio binaries in such
environments. Considerable work has been done on this problem,
including the pioneering analysis of Heggie (1975) and the
numerical experiments of Hills (1975) and Hut \& Bahcall (1983),
and more recently simulations directed at specific
problems in globular clusters \cite{mb95,md95,sp93,sp95}.
Relatively little work has been done on the problem of very
high mass ratio systems appropriate to planetary encounters,
Sigurdsson (1992) considered the basic problem, Laughlin \& Adams
calculated mean disruption rates for open clusters (1998),
and Bonnell \& Kroupa (1998) looked at the issue in the context
of planet formation. Hills (1984) considered close encounters between a 
star-planet system and an intruder star. Hills \& Dissly (1989) extended
this work to consider a range of impact velocities and intruder masses.
For the M4 system, considerable effort has been
made to simulate its possible dynamical history and the implications
of the existence of such a system \cite{si93,si95,jo97,fo00}.

47 Tucanae is a massive and quite dense cluster.
The central density is about $1.5\times 10^5\ {\rm pc^{-3}}$
with central velocity dispersion of $\sim 12 {\rm km\, s^{-1}}$.
The core radius is about 0.5 pc and the half-mass radius about 3.9 pc
\cite{dj93,gu00}. At the half mass radius the number density of stars
is about a factor of ten lower than in the core.

\section{Encounter Timescales}

Encounters between two stars will be extremely rare in the low-density
environment of the solar neighbourhood. However, in the cores of globular
clusters number densities are sufficiently high 
($\sim 10^5$ stars/pc$^3$ in some systems) that
encounter timescales can be comparable, or even less than, the 
age of the universe. In other words, a large fraction of the stars
in these systems will have suffered from at least one close encounter
or collision in their lifetime.

The cross section for two stars, having a relative velocity at infinity
of $V_\infty$, to pass within a distance $R_{\rm min}$ is given by

\begin{equation}
\sigma = \pi R_{\rm min}^2 \left( 1 + {V^2 \over V_\infty^2} \right)
\end{equation}

\noindent where $V$ is the relative velocity of the two 
stars at closest approach in a parabolic encounter ({\it i.e.\ } 
$V^2 = 2 G (M_1 + M_2)/R_{\rm min}$).
The second term is due to the attractive gravitational force, and
is referred to as gravitational focussing.
If $V \gg V_\infty$
as will be the case in systems with low velocity dispersions, such as
globular clusters, $\sigma \propto R_{\rm min}$.

One may estimate the timescale for a given star to undergo an encounter,
$\tau_{\rm enc} = 1/n \sigma v$. For clusters with low velocity dispersions,
we thus obtain

\begin{equation}
\tau_{\rm enc} = 7 \times 10^{10} {\rm yr} \left( {10^5pc^{-3} \over
n } \right) \left( { v_\infty \over 10{\rm km/s}} \right) 
\left( { R_\odot \over R_{\rm min} } \right) \left( { M_\odot \over
M } \right) 
\end{equation} 

\noindent where $n$ is the number density of single stars of mass $M$.

We may estimate the timescale for an encounter between a star-planet system and a 
single star, in a similar manner where now $R_{\rm min} \simeq d$, 
where $d$ is the semi-major axis of the planet's orbit. The encounter
timescale for such a system may therefore be relatively short as the semi-major
axis can greatly exceed stellar radii.
For example, a planetary system with $d \sim 1AU (\equiv 216
R_\odot)$, will have an encounter timescale $\tau_{\rm enc} 
\lo 10^{9}$ years in the core of a dense globular cluster.

Encounters between single stars and binary systems
will lead to the break up of the binaries when the kinetic energy of 
the incoming star exceeds the binding energy of the binary. This will be
the case when $V_\infty \geq V_{\rm c}$, where 

\begin{equation}
V_c^2={ G M_1 M_2(M_1+M_2+M_3) \over M_3(M_1+M_2)d }
\end{equation}

\noindent where $M_1$ is the mass of the
primary, $M_2$ the mass of the secondary and $M_3$ is the mass of the
incoming, single star. Binaries vulnerable to break up are often referred
to as being {\it soft} \cite{he75,hb83}.
In the case of encounters between single stars
and a star-planet system, the latter will be invariably soft as $M_2 \ll M_1$ and $M_3$.
We would therefore expect the planetary system to be broken up 
with some high probability, when an intruding
star passed within a distance $d$ of the planetary-system star (but see Sigurdsson 1993).
In addition, more distant encounters may leave the planets in 
somewhat eccentric orbits, as will be discussed later.
The importance of encounters between a planetary system and binaries is
a function of the binary fraction within the cluster, which is not well known.

\section{Numerical methods}

For encounters involving a single intruding star, we used a code based
on the one discussed in Davies, Benz \& Hills (1993).
For encounters involving binaries (\ie four objects), we used a code based
on the one discussed in Bacon, Sigurdsson \& Davies (1996).

The initial conditions for the scatterings were set following
the method of Hut \& Bahcall (1983, see also Sigurdsson \& Phinney 1993).
With two binaries, we have additional parameters from the relative
phase of the second binary, the orientation of the plane of the second
binary and the second binary mass ratio, semi--major axis
and eccentricity. We drew the binary parameters
by Monte Carlo selection uniformly over the phase variables.
The relative velocity at infinity of the centres--of--mass of the two
binaries, $v_{\infty}$ was chosen uniformly on the interval allowed.
The initial binary eccentricities, $e_{1,2}$ were zero for all encounters,
previous calculations indicate the cross--sections of
interest are not sensitive to the binaries' eccentricities.

For encounters involving single stars, the equations of motion are
rescaled using the Heggie (1972) quasi regularisation and are integrated
using the variable-order, variable-stepsize Shampine-Gordon integrator (1975).
For encounters involving binaries, 
a Bulirsch--Stoer variable step integrator
with KS--chain regularisation (Aarseth 1984, Mikkola 1983, 1984a,b)
was used for integrating the motion of the particles.

\section{Results}

\begin{table}
\begin{tabular}{lllll} \hline\hline
\noalign{\medskip}
$M_{2}/{\rm M}_\odot$ & $\alpha$ & $V_{\rm orb}$ & $V_\infty$ & $V_{\rm c}$ \\
\noalign{\medskip}
\hline
\noalign{\medskip}
0.1   & 1.0  &  31.24 & 31.24 & 13.01 \\ 
0.01  & 1.0  &  29.94 & 29.94 & 4.20  \\
0.001 & 1.0  &  29.80 & 29.80 & 1.33  \\
\noalign{\medskip}
0.01  & 3.0  &  29.94 & 89.82 & 4.20  \\
0.01  & 0.55 &  29.94 & 16.46 & 4.20  \\  
0.01  & 0.30 &  29.94 & 8.98  & 4.20  \\ 
0.01  & 0.17 &  29.94 & 5.09  & 4.20  \\
0.01  & 0.10 &  29.94 & 2.99  & 4.20  \\
\noalign{\medskip}
0.001 & 0.30 &  29.80 & 8.94  & 1.33  \\
\noalign{\medskip}
\hline
\end{tabular}
\caption{A list of the binary-single star encounters studied. $\alpha =
V_\infty/V_{\rm orb}$ and $V_{\rm c}$ is the critical velocity given by
equation (3). $M_2$ is the planet mass. All velocities are given in km/s.}
\end{table}

\begin{table}
\begin{tabular}{lllll} \hline\hline
\noalign{\medskip}
$M_{2}/\msun$ & $\alpha$ & $V_{\rm orb}$ & $V_\infty$ & $V_{\rm c}$ \\
\noalign{\medskip}
\hline
\noalign{\medskip}
0.01  & 3.0  &  29.94 & 89.82 & 36.54  \\
0.01  & 1.0  &  29.94 & 29.94 & 36.54  \\
0.01  & 0.30 &  29.94 & 8.98  & 36.54  \\ 
0.01  & 0.10 &  29.94 & 2.99  & 36.54  \\ 
\noalign{\medskip}
\hline
\end{tabular}
\caption{A list of the binary-binary encounters studied. $\alpha =
V_\infty/V_{\rm orb}$ and $V_{\rm c}$ is the critical velocity given by
equation (4). $M_2$ is the planet mass. All velocities are given in km/s.}
\end{table}

\begin{figure}
\psboxto(\hsize;0cm){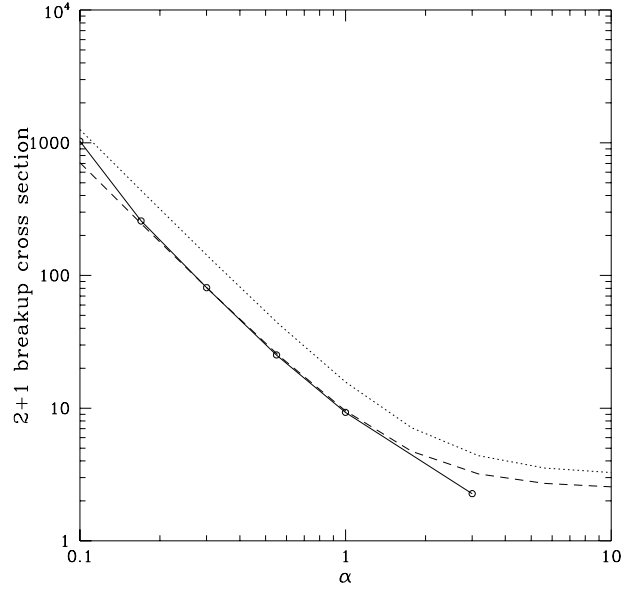}
\caption{The cross section for breakup of the planet-star system as a function
of $\alpha= V_\infty/V_{\rm orb}$
(solid line), and for the intruding star to pass within the orbital separation
of the planet-star system (dotted line). Both cross sections are given in units
of the binary separation (ie the cross section of a target having a radius
equal to the binary separation is $\pi$). 
The dashed line is the breakup cross section for $\alpha=0.3$ rescaled
assuming that the change in cross section is purely an effect 
of the differing degree
of gravitational focussing ocurring for different values of $\alpha$.}
\end{figure}

\begin{figure}
\psboxto(\hsize;0cm){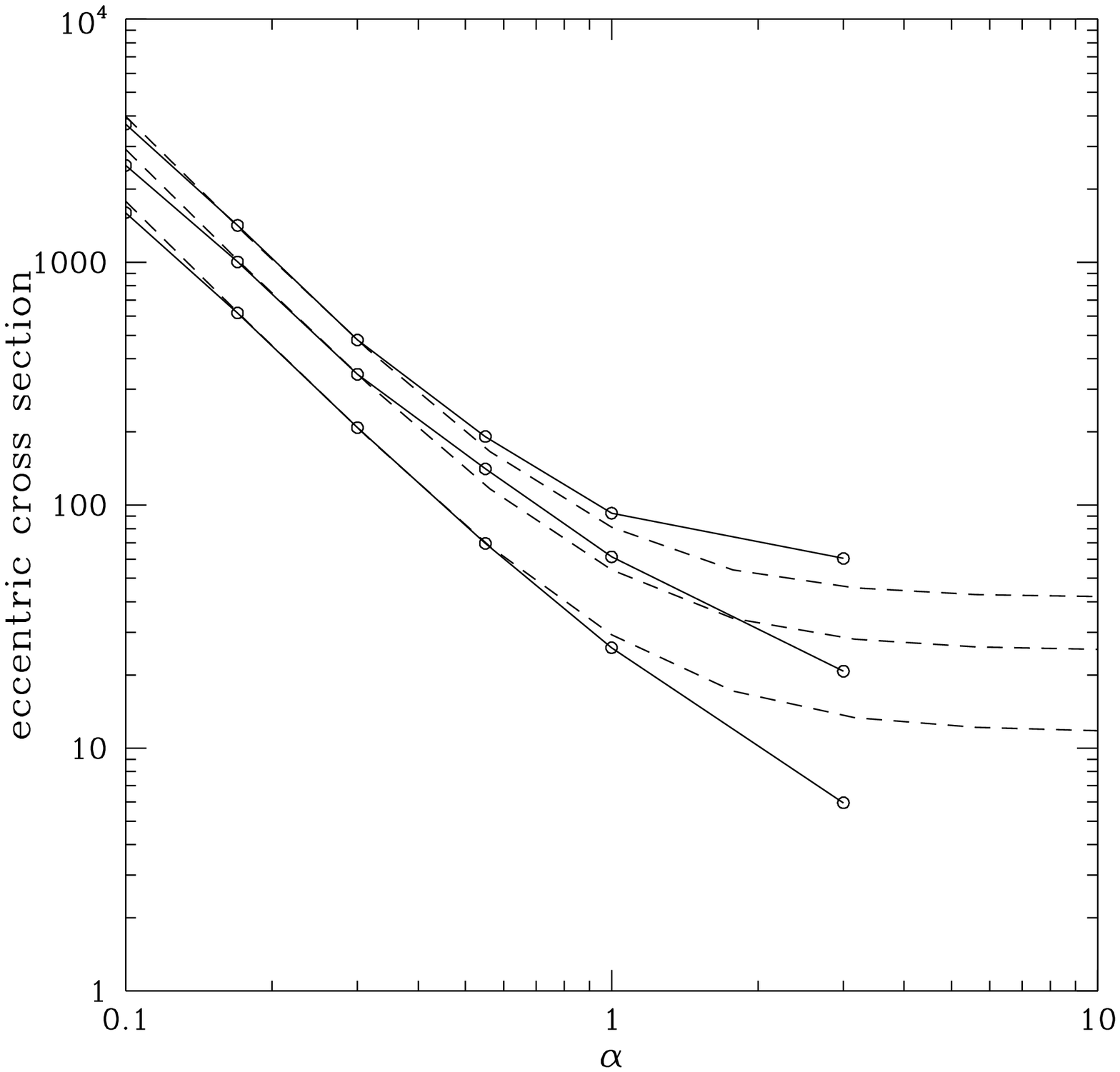}
\caption{The cross section for the planet to be left in an eccentric 
orbit as a function of $\alpha= V_\infty/V_{\rm orb}$. Cross sections 
are for systems with eccentricity $e> 0.03$,
0.1 and 0.3 (top to bottom).
All cross sections are given in units
of the binary separation as in Figure 1. 
The dashed line is the breakup cross section for $\alpha=0.3$ rescaled
assuming that the change in cross section is purely an effect 
of the differing degree
of gravitational focussing ocurring for different values of $\alpha$.}
\end{figure}

We considered encounters between a star-planet system and either a single star
or a binary. The star-planet system contained a single 
planet in a circular orbit. 
We consider encounters for various values of $\alpha$, 
where $\alpha=V_\infty/V_{\rm orb}$.
These will correspond to different values of $V_\infty$ 
as a function of the radius
of the planetary orbit, $d$, and thus $V_{\rm orb}$. The encounters considered 
are listed in Tables 1 and 2. In both tables, the values of $V_{\rm orb}$ and
$V_\infty$ are given in km/s assuming $d=1$ AU. In Table 1, $V_{\rm c}$ is given
by equation (3) whereas in Table 2, the velocity at which the total energy is
zero, $V_{\rm c}$ is given by

\begin{equation}
V_c^2 = G\mu \biggl (  { {M_1M_2}\over {d_1} }+{ {M_3M_4}\over {d_2} } \biggr )
\end{equation}

\noindent where $\mu = M_T/(M_1+M_2)(M_3+M_4)$, $M_T = M_1 + M_2 + M_3 + M_4$,
is the binaries reduced mass, and $d_1, d_2$ are the semi--major axis of
the binaries containing masses $M_{1,2}, M_{3,4}$ respectively.

Note for all simulations, $M_1=M_3=M_4=1.0 {\rm M}_\odot$, and $d_1=d_2$.
The turnoff mass for 47 Tuc is slightly lower than $1\, M_{\odot}$, probably
$0.85 \pm 0.05 M_{\odot}$, and the true encounter rates scale 
to be correspondingly
slightly lower.

For each value of $\alpha$, we simulated encounters 
for a range of impact parameters, ensuring
that we considered sufficient large impact 
parameters to include all relevant interactions.
For each set of encounters we then computed 
the cross sections for the planetary system
to be disrupted
(\ie the planet does not remain in orbit around {\it either} star), 
and for the planet's orbit to be perturbed to eccentricities, $e \geq 0.03$,
0.1, and 0.3.

The cross section for the breakup of the planetary system via encounters with a single star
is shown in Figure 1 as a function of $\alpha$. Also shown is the cross section for
the two stars to pass within a distance, 
$R_{\rm min} \leq d$ which was calculated using equation (1). 
Also shown is the breakup cross section for $\alpha=0.3$ rescaled
assuming that the change in cross section is purely an effect 
of the differing degree
of gravitational focussing ocurring for different values of $\alpha$.

The cross section for the breakup of the planetary system 
via encounters with binary stars
shows a similar form to that seen in Figure 1, 
except the value is increased by a factor $\sim$ 2.

We consider now the cross sections for the production of eccentric 
planetary systems. These are
shown in Figure 2 for encounters with single stars. 
The cross sections are larger than
the equivalent breakup cross section. The cross section 
for producing a planetary system
with eccentricity $e > 0.03$ being $\sim$ 3--5 $\sigma_{\rm breakup}$ 
(see also Heggie \& Rasio 1996).
An important point to note here is that the systems with 
non-zero eccentricities can have
semi-major axes larger than the initial circular 
orbit of the planet, and in any case
the maximum separation (\ie $d_{\rm new} (1+e)$) 
is certainly larger. A planet left in such
an eccentric orbit is therefore a larger target 
for a subsequent breakup. We estimate that
this will double the effective breakup cross section for 
planets with initial separations
$d \sim$ 0.1 AU. Systems left in extremely eccentric 
orbits tend to have equivalently 
large semi-major axes, this being the limit 
of sytems that just failed to be broken up.
The number of eccentric systems where 
circularisation occurs will be small and limited
to systems where the initial separation $d \lo 0.1$ AU.

If we consider that $V_\infty$ is a constant value for all sets of encounters,
then the various values of $\alpha$ correspond to different 
values of orbital separation $d$. As $\alpha = V_\infty/V_{\rm orb}$ 
and $V_{\rm orb} \propto \sqrt{1/d}$, $d \propto
\alpha^2$. Considering $V_\infty = 9$ km/s, inspection of Table 1 reveals that
$\alpha = 0.3$ corresponds to $d = 1$ AU. 
rescaling the other cross sections for the same
value of $V_\infty$ we produce Figures 3 and 4, 
where now the cross sections are given
in units of (AU)$^2$. These may be converted to timescales via

\begin{equation}
\tau = 4.16 \times 10^{15} \ {\rm yr} \ \left( { 1 \over n} \right) \left( {1 \over
\sigma } \right)
\end{equation}

\noindent where $n$ is the number density of stars (or binaries)/pc$^3$, 
and $\sigma$ is the cross sections in units of (AU)$^2$. 
In Table 3 we give the timescales in years for systems to be 
broken up or left in eccentric orbits for both binary--single
and binary--binary encounters, assuming a number density, $n = 10^5$ 
stars/pc$^3$, and $V_\infty = 9$ km/s.
We will consider the implications for the survival, 
or otherwise, of planetary systems in stellar clusters 
in the discussion section. 

\begin{figure}
\psboxto(\hsize;0cm){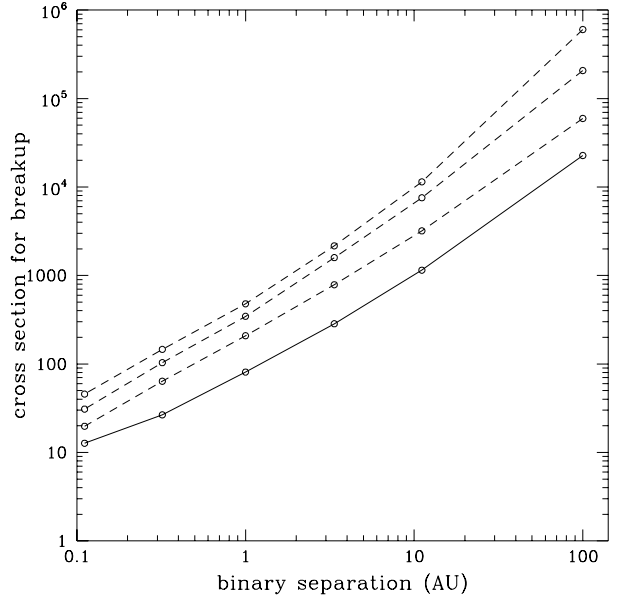}
\caption{The cross section for breakup of the planet-star system 
from encounters with single stars as a function
of separation in AU (solid line), and for the planet to be
left in an eccentric orbit with eccentricity, 
$e >$ 0.03, 0.1, 0.3 (dashed lines,
from top to bottom).}
\end{figure}

\begin{figure}
\psboxto(\hsize;0cm){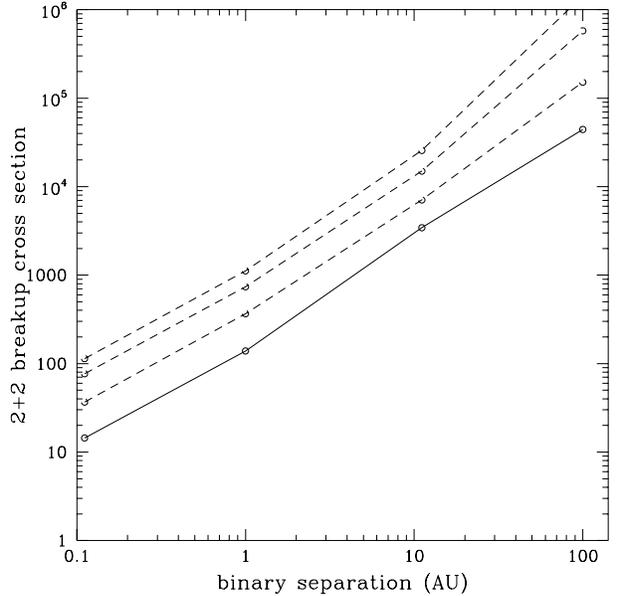}
\caption{The cross section for breakup of the planet-star system 
from encounters with binary stars as a function
of separation in AU (solid line), and for the planet to be
left in an eccentric orbit with eccentricity, 
$e >$ 0.03, 0.1, 0.3 (dashed lines,
from top to bottom).}
\end{figure}

\subsection{The importance of collisions}

Thus far we have only considered exchanges and breakups as mechanisms to destroy planetary
systems. For the very tightest systems collisions between the planet and a star or between
two stars will also play a role. If two stars collide and merge, the planet will likely
be swallowed up by the expanding merged object, assuming it remainds bound to the
collision product in the first place.

We computed the cross sections for physical collisions or very 
close encounters between objects during an interaction. For encounters 
involving a single star, we find that mergers
are unimportant except for encounters involving very tight 
planetary systems (\ie $d\lo$ 0.1
AU) where $\sigma_{\rm col} \sim 0.5 \sigma_{\rm breakup}$. 
For encounters involving
binary stars, we find that the cross section for mergers can be larger than the 
breakup cross section for $d \lo 0.5$ AU. For such tight planetary systems, 
we find
that $\sigma_{\rm col} \sim 150$ in units of (AU)$^2$.

\section{Discussion}

\begin{table}
\begin{tabular}{lllll} \hline\hline
\noalign{\medskip}
$d/{\rm AU}$ &$\tau_{\rm breakup}$ &  $e>0.3$  &  $e>0.1$  & $ e> 0.03$ \\
\noalign{\medskip}
\hline
\noalign{\medskip}
\noalign{binary--single encounters}
\noalign{\medskip}
  1.0000& 5.2E+08& 2.0E+08& 1.2E+08& 8.7E+07 \\
  0.1000& 5.2E+09& 2.0E+09& 1.2E+09& 8.7E+08 \\
  0.0500& 1.0E+10& 4.0E+09& 2.4E+09& 1.7E+09 \\
  0.0300& 1.7E+10& 6.7E+09& 4.0E+09& 2.9E+09 \\
\noalign{\medskip}
\noalign{binary--binary encounters}
\noalign{\medskip}
  1.0000& 3.0E+08& 1.1E+08& 5.6E+07& 3.7E+07 \\
  0.1000& 3.0E+09& 1.1E+09& 5.6E+08& 3.7E+08 \\
  0.0500& 6.0E+09& 2.7E+09& 1.1E+09& 7.4E+08 \\
  0.0300& 1.0E+10& 3.8E+09& 1.9E+09& 1.2E+09 \\
\noalign{\medskip}
\hline
\end{tabular}
\caption{Timescales (in years) for systems to be 
broken up or left in eccentric orbits for both binary--single
and binary--binary encounters, assuming a number density, $n = 10^5$ 
stars/pc$^3$, and $V_\infty = 9$ km/s.}
\end{table}

Combining the cross sections for breakups, 
significantly increasing system size (by
making it eccentric), and collisions, we estimate that the effective cross
section for the destruction of tight planetary systems 
$\sigma_{\rm destroy} \sim 30$
for encounters involving single stars, and $\sigma_{\rm destroy} \sim 150$ for
encounters involving binary stars. Assuming the core binary fraction in 
47 Tuc to
be 0.1 -- 0.2, we therefore conclude that effective destruction cross section 
in  the core is $\sigma_{\rm destroy} \sim 40 - 50$,
for a planet in 0.4 AU orbit. The data from the HST project will
give the fraction of high mass-ratio main sequence binaries
in the same part of the cluster surveyed for planets.

Using equation (5), we are able to compute the likely timescale 
for the destruction of a planetary system, or the time required 
to place the planet on an eccentric orbit.
Assuming that a particular star-planet system is located within 
the core of 47 Tuc, where $n \sim 10^5$ stars/pc$^3$, 
we see that $\tau_{\rm breakup} \sim 4 \times 10^{10}/
\sigma_{\rm destroy}$ years. We therefore require $\sigma_{\rm destroy} \go 4$ 
which is easily satisfied for most semi-major axes. A planet with an orbital
separation $d \sim 5$ AU would be broken up in $\sim 10^8$ years.
By comparison, a planet orbiting a star at the cluster half-mass 
radius will be somewhat less vulnerable. Here the number density 
$n \sim 10^4$ stars/pc$^3$ 
hence we require $\sigma_{\rm destroy} \go 40$;
only those systems where $d \go$ 0.3 AU are likely to be destroyed.
The critical value of $d$ will clearly depend on the exact value
of the number density of stars, $n$.
Note that Sigurdsson (1992) assumed a stellar density $\sim 3$ times
smaller for 47 Tuc and derived a correspondingly longer time scale for
breakup in the core.
Additional uncertainty is introduced by the time evolution of the
47 Tuc stellar density profile. If the cluster is evolving through
normal relaxation, then it was substantially less dense in the centre
a few billion years ago. However, there is some suggestion that the
cluster may already have evolved through a dense core phase \cite{si00}
in which case rate of destruction of tightly bound planetary systems may 
have been an order of magnitude higher in the recent past.
In either case, theory predicts little or no evolution in the stellar density
near the half--mass radius, and the WF/PC2 field extends out from the core
out almost to the half--mass radius. On the other hand, there are more stars
to be oberved in the denser parts of the cluster and the prior expectation
therefore that most planets should have been found in the densest part
of the observed field, assuming no dynamical disruption took place. 


It should also be noted that simulations of the evolution of 
multiple planetary systems suggest that a relatively 
slight perturbation in the eccentricity of one or more planets
may lead to a radical rearrangement of the system in a short 
timescale \cite{qu92,mh99}
Thus we note that encounters that perturb the planets may play 
a role in the evolution of planetary systems, even in relatively 
low-density short-lived stellar clusters.
For example, if $n \sim 10^3$ stars/pc$^3$, $\tau_{\rm enc} \sim 10^7$ years, 
for 
$d \go 30$ AU.

In terms of the current survey in 47 Tuc, it would seem possible that all
planetary systems have been destroyed if all the stars in the survey were 
located in the highest-density regions. Given that some of the stars observed
are in regions with an order of magnitude smaller stellar density, 
it is difficult to explain away all the expected detections.
However, there are fewer stars observed in the 
low-density regions of the cluster,
so if the intrinsic specific density of 51 Peg 
like systems is somewhat lower than
in the solar neighbourhood, 
then we might have expected more like 5-10 detections
from observing 34,000 stars, and 75\% of those in the densest regions.
In order for direct dynamical disruption to explain the absence of 
51 Peg like systems, one still requires either a past denser phase for the core,
or a very long time spent at the current density.
The minimum average density for the
environment of a given planetary system required for its destruction
being $\sim 10^4$ stars/pc$^3$. From our calculations here, we conclude that
tight planetary systems (\ie $d \lo 0.1$ AU) will definitely survive
in less-dense clusters
where subsequent surveys should be conducted.
An ideal target is NGC~6352. It has core density of about $10^3$ stars/pc$^3$,
is at an estimated distance of 6.1 kpc (compared to 4.6 kpc for 47 Tuc)
and has metallicity [Fe/H] = -0.51 (compared with -0.71 for 47 Tuc).
With the Advanced Camera on the {\it Hubble Space Telescope}, a considerably
wider field can be monitored, and another survey for transits carried out.

\section*{ACKNOWLEDGEMENTS}
MBD gratefully acknowledges the support of a URF from the Royal Society. 
SS gratefully acknowledges support from NASA through grants STSCI 
GO-7307 and GO-8267.


\begin{thebibliography}{}
\bibitem[]{} Aarseth S., 1984, in Goodman J., Hut P., eds., Proc. IAU Symp. 113,
Dynamics of Star Clusters, Reidel, Dordrecht, p. 251 
\bibitem[Backer et al 1993]{ba93} Backer, D.C., Foster, R.S. \& Sallmen, S., 1993, Nature, 365, 817
\bibitem[Bacon et al 1996]{ba96} Bacon, D., Sigurdsson, S. \& Davies, M.B., 1996, MN, 281, 830
\bibitem[Bonnell \& Kroupa 1998]{bo98} Bonnell, I. \& Kroupa, P., 1998, 
in ``The Orion Complex Revisited'', eds McCaughrean \& Burkert, 
PASP Conf Proc., in press (astro-ph/9802306)
\bibitem[Brown et al 2000]{br00} Brown, T., et al, 2000 (in preparation)
\bibitem[Charbonneau et al 2000]{ch00} Charbonneau, D., 
Brown, T.M., Latham, D.W. \& Mayor, M., 2000, ApJL, 529, L45
\bibitem[]{}  Davies M.B., Benz W., Hills J.G., 1993, ApJ, 411, 285
\bibitem[Davies \& Benz 1995]{mb95}  Davies M.B., Benz W., 1995, MN, 276, 876
\bibitem[Davies 1995]{md95}  Davies M.B., 1995, MN, 276, 887
\bibitem[Djorgovski 1993]{dj93} Djorgovski, G., 1993, in PASP Conf Proc Vol. 50,
Structure and Dynamics of Globular Clusters, eds. Djorgovski, S.G., Meylan G.
\bibitem[Ford et al 2000]{fo00} Ford, E.B., Joshi, 
K.J., Rasio, F.A. \& Zbarsky, B., 2000, ApJ, 528, 336
\bibitem[Gilliland et al 2000]{gi00} Gilliland, R.L., et al, 2000 
(in preparation)
\bibitem[Heggie 1972]{he72} Heggie, D.\ C., 1972, 
in The Gravitational N-body Problem,
ed. M. Lecar (Dordrecht: Reidel), 148
\bibitem[Heggie 1975]{he75} Heggie, D.C., 1975, MN, 173, 729
\bibitem[Heggie \& Rasio 1996]{hr96} Heggie, D.C. \& Rasio, F.A., 
1996, MN, 282, 1064
\bibitem[Hills 1975]{hi75} Hills, J.G., 1975, AJ, 80, 809
\bibitem[]{} Hills, J.G., 1984, AJ, 89, 1559
\bibitem[]{} Hills, J.G., Dissly, R.W., 1989, AJ, 98, 1069
\bibitem[Howell \& Guhathakurta 2000]{gu00} Howell, J.H. \& 
Guhathakurta, R., 2000, preprint
\bibitem[Hut \& Bahcall 1983]{hb83}  Hut, P., \& Bahcall, J.N., 
1983, ApJ, 268, 319
\bibitem[Joshi \& Rasio 1997]{jo97} Joshi, K.J. \& Rasio, F.A., 
1997, ApJ, 488, 901
\bibitem[Laughlin \& Adams 1998]{la98} Laughlin, G., \& Adams, F.C., 
1998, ApJL, 508, L171
\bibitem[Lyne et al 1988]{ly88} Lyne, A.G., Biggs, J.D., Brinklow, A., 
McKenna, J. \& Ashworth, M., Nature, 332, 45
\bibitem[Marcy \& Butler 1996]{ma96} Marcy, G.W. \& Butler, R.P., 
1996, ApJL, 464, L147
\bibitem[Mayor \& Queloz 1995]{ma95} Mayor, M. \& Queloz, D., 
1995, Nature, 378, 355
\bibitem[]{} Mikkola S., 1983, MN, 203, 1107
\bibitem[]{} Mikkola S., 1984a, MN, 207, 115
\bibitem[]{} Mikkola S., 1984b, MN, 208, 75
\bibitem[Murray \& Holman 1999]{mh99} Murray, N. \& Holman, M., 
1999, Science, 283, 1877
\bibitem[Quinlan 1992]{qu92} Quinlan, G.D., 1992, in ``Chaos, Resonance, 
and Collective Dynamical Phenomena in the Solar System'', IAU Symp. 152, 
p. 25 (Kluwer - Dordrecht)
\bibitem[]{} Shampine, L.\ F., Gordon, M.\ K., 1975, in Computer
Solutions of Ordinary Differential Equations: 
The Initial Value Problem (New York: Freeman)
\bibitem[Sigurdsson 1992]{si92} Sigurdsson, S., 1992, ApJL, 399, L95
\bibitem[Sigurdsson 1993]{si93} Sigurdsson, S., 1993, ApJL, 415, L43
\bibitem[Sigurdsson \& Phinney 1993]{sp93} Sigurdsson, S.,  
Phinney, E.\ S., 1993, ApJ, 415, 631
\bibitem[Sigurdsson \& Phinney 1995]{sp95} Sigurdsson, S., 
Phinney, E.\ S., 1995, ApJS, 99, 609
\bibitem[Sigurdsson 1995]{si95} Sigurdsson, S., 1995, ApJ, 452, 323
\bibitem[Sills et al 2000]{si00} Sills, A., Bailyn, C.D., 
Edmonds, P.D. \& Gilliland, R.L., 2000, ApJ, 535, 298
\bibitem[Thorsett et al 1993]{th93} Thorsett, S.E., Arzoumanian, Z. 
\& Taylor, J.H., 1993, ApJL, 412, L33
\bibitem[Thorsett et al 1999]{th99} Thorsett, S.E., Arzoumanian, Z., 
Camilo, F. \& Lyne, A.G., 1999, ApJ, 532, 763
\bibitem[Wolszczan \& Frail 1992]{wo92} Wolszczan, A. \& Frail, 
D.A., 1992, Nature, 355, 145
\end{thebibliography}
\end{document}